\documentclass[useAMS,usenatbib,fleqn]{mn2e}
\usepackage{amsmath,amssymb,mathrsfs,graphicx,float,latexsym}
\usepackage{epstopdf,ragged2e}
\usepackage{hyperref}

\newcommand{\tH}{t_{\text{H}}}
\newcommand{\EF}{E_{\text{FRB}}}
\newcommand{\Eq}{E_{\text{quake}}}
\newcommand{\sm}{\sigma_{\text{max}}}
\newcommand{\bb}{\boldsymbol{B}}

\def\apj{{ApJ}}

\def\apjs{{The Astrophysical Journal Supplement}}
\def\apjl{{ApJL}}

\def\aap{{A\&A}}

\def\mnras{{MNRAS}}

\def\pasj{{Publications of the Astronomical Society of Japan}}

\def\nat{{Nature}}
\def\aapr{{The Astronomy and Astrophysics Review}}

\def\prd{{Physical Review D}}

\def\04a{{2004 a}}
\def\04b{{2004 b}}

\title[Magnetars with fracturing crusts as FRB repeaters]{Young magnetars with fracturing crusts as fast radio burst repeaters}
\author[A. G.~Suvorov and K. D.~Kokkotas]{A. G.~Suvorov\thanks{E-mail: arthur.suvorov@tat.uni-tuebingen.de} and K. D.~Kokkotas\\Theoretical Astrophysics, Eberhard Karls University of T{\"u}bingen, T{\"u}bingen, D-72076, Germany}
\begin{document}

\date{Accepted ?. Received ?; in original form ?}

\pagerange{\pageref{firstpage}--\pageref{lastpage}} \pubyear{?}

\maketitle
\label{firstpage}
%While the physical origin of these phenomena is unknown, the discovery of repeating sources suggests that active, compact objects may be responsible. matching the predicted age of the object within FRB 121102. In particular, pulse energies can be related to the depths and lengths of the fractures, which may occur sporadically over the stellar surface if the magnetic field is topologically complicated.

\begin{abstract}

\noindent{Fast radio bursts are millisecond-duration radio pulses of extragalactic origin. A recent statistical analysis has found that the burst energetics of the repeating source FRB 121102 follow a power-law, with an exponent that is curiously consistent with the Gutenberg-Richter law for earthquakes. This hints that repeat-bursters may be compact objects undergoing violent tectonic activity. For young magnetars, possessing crustal magnetic fields which are both strong ($B \gtrsim 10^{15}$ G) and highly multipolar, Hall drift can instigate significant field rearrangements even on $\lesssim$ century long timescales. This reconfiguration generates zones of magnetic stress throughout the outer layers of the star, potentially strong enough to facilitate frequent crustal failures. In this paper, assuming a quake scenario, we show how the crustal field evolution, which determines the resulting fracture geometries, can be tied to burst properties. Highly anisotropic stresses are generated by the rapid evolution of multipolar fields, implying that small, localised cracks can occur sporadically throughout the crust during the Hall evolution. Each of these shallow fractures may release bursts of energy, consistent in magnitude with those seen in the repeating sources FRB 121102 and FRB 180814.J0422+73.
}

%the crust yielding may appear stochastic, explaining the statistical curiosity that the energetics and waiting times of FRB 121102 are seemingly uncorrelated.

\end{abstract}


\section{Introduction}

Fast radio bursts\footnote{{A catalogue of observed FRBs is maintained at http://frbcat.org/} \citep{petroff16}.} (FRBs) are short (duration $\sim$ ms) but intense (flux $\lesssim$ Jy) flashes, generally believed to be of extragalactic origin due to their high dispersion measures, which appear in the GHz-band \citep{lori07,keane12,thornton13,ravi15,gourd18,zhang19,petroff19}. The physical mechanisms driving these phenomena are unknown, though, considering the timescale and energy requirements, most proposals involve coherent emission associated with disrupted compact objects, such as neutron stars (NSs) \citep{totani13,lyu14,pen15,wang18} or black holes \citep{ravi14,ming15,zhang16,barrau18}. Moreover, the existence of repeating sources, of which two have now been discovered, FRB 121102 \citep{spitler14,tend17,scholz17,gajjar18} and FRB 180814.J0422+73 \citep{chime19,wang19,yang19}, suggests that there is at least a subclass of FRBs resulting from transient outbursts of a young object \citep{caleb18,palan18}. For FRB 121102, the constraints coming from the dispersion measure, GHz free-free optical depth, and the size of the quiescent source indicate that the progenitor is $\gtrsim 30-100$ yrs old \citep{kash16,kash17,bower17,metzger17} [see fig. 1 of \cite{marg18}].

%(accounting for photo-ionization by the rotationally-powered magnetar nebula) %through the expanding supernova ejecta Somefraction of the seismic energy will be damped, heating the crus %Sporadic releases of energy ae therefore possible if many local patches of stress accumulate over the crustal surface thereby causing multiple, small fractures, each of which contribute to the overall transient activity.

Owing to the multitude of theorised FRB progenitors [see \cite{katz18,platts18} for {recent} reviews], it is clear that we must turn to observational clues to try and discriminate between different scenarios. \cite{wang18} found that the burst energetics of FRB 121102 follow a distribution similiar to that of the \cite{gutenberg56} law associated with earthquakes. This statistical similarity suggests that NS tectonic activity and eventual crustquakes may be, at least partially, responsible for [amongst other things, e.g. ($\text{anti-}$)glitches \citep{eps00,mast15}, giant flares \citep{thomp02,col11,zink12}, or quasi-periodic oscillations \citep{thomp17}] the repeated FRBs [see also \cite{zhang18}]. {Furthermore, low-energy $\gamma$-ray flashes \citep{cheng96}, giant flares \citep{xu06}, and short bursts \citep{wad19} from mature soft-gamma repeaters (SGRs) also follow earthquake-like statistics, evidencing a connection between magnetar outbursts and crustal activity \citep{wang17}.}

On the other hand, \cite{li19} found no correlation between the waiting times and energetics from the $170$ pulses thus far observed in FRB 121102, and argued that this implies that the mechanism responsible for repeating sources is unlikely to be intrinsic to the star; one might expect that, when more time has elapsed between subsequent bursts, the energetics should increase. However, the extent to which a crust is susceptible to shear stresses depends on its molecular properties \citep{horo09,chug10,hoff12,chug18}, and complicated fracture geometries may result if highly anisotropic stresses are exerted \citep{link00}. While the formation of small-scale voids or fractures in older NSs may be restricted by the high pressure-to-shear-modulus ratio \citep{jones03,horo09} [see also \cite{levin12}], a very young, more pliable crust may be susceptible to local fracturing. In particular, sporadic energy releases via minor quakes are possible if many local patches of stress accumulate within the crustal layers, thereby causing multiple, small fractures, each of which contribute to the overall transient activity. An evolving magnetic field with a complicated topological structure, for example, would be expected to strain the crust in a highly anisotropic manner \citep{lander15,lan19}. 

Various mechanisms can give rise to strong, multipolar magnetic fields within proto-NSs. In a core-collapse scenario, the characteristic field strength of the collapsed star is set, to first-order, by magnetic flux conservation ($B \lesssim 10^{13} \text{ G}$). Within seconds after collapse, strong core-surface differential rotation stretches field lines and generates strong, mixed poloidal-toroidal fields through wind-up \citep{jank89,spruit99,braith06}. The differential rotation combined with turbulent convection drives an $\alpha-\omega$ dynamo, which may generate strongly multipolar fields with characteristic strengths of the order $\sim 10^{16} \text{ G}$ \citep{dunc1,dunc2,miralles02}. The field may then be further amplified by the magnetorotational instability \citep{shibata06,sawai13}, shockwave instabilities from the core bounce \citep{endeve12}, or electric currents generated from chiral imbalances between charged fermions in the plasma \citep{zanna18}. For a star born following a NS-NS merger, the Kelvin-Helmholtz instability, generated at the shear layer between the progenitor stars, can amplify the field up to $\lesssim 10^{17} \text{ G}$ \citep{price06,giac15,gourg18}.

Even after stable stratification $\gtrsim 10$ s after formation, the magnetic field may continue to evolve rapidly through superfluid turbulence \citep{ferrario10}, which drives the circulation of charged particles via mutual friction and entrainment \citep{peralta05}, ambipolar diffusion, associated with direct Urca processes in stars with cooler $(T \lesssim 10^{9} \text{ K})$ and superconducting cores \citep{pass17}, or, in stars with high surface electron temperatures $T_{e} \sim 10^{7} \text{ K}$, from thermoelectric instabilities \citep{urpin86,gepp17}. The efficacy of each of these mechanisms is still not well understood, though certainly depends critically on the properties of the progenitor(s), such as their mass, angular momentum, magnetic field strength, and metallicity \citep{herant92,heger03,scheid10,nakamura14,vart18}. Under some (possibly rare) conditions therefore, these effects might combine to yield a particularly tangled, `turbulent' (i.e. strongly multipolar, and with non-negligible toroidal component) magnetic field shortly after birth, as supported by core-collapse \citep{ober17,ober18} and NS-NS merger simulations \citep{giac15,ciolfi19}. Furthermore, measurements of phase-resolved cyclotron absorption lines reveal that some magnetars possess local magnetic structures orders of magnitude stronger than their dipole-spindown fields \citep{sanwal02,tiengo13}.

%In either case, the magnetic field formed is likely to be outof equilibrium and unstable directly after the collapse, raisingthe  question  how  much  of  the  magnetic  field  can  survive  inthe  subsequent evolution  when  the  star  is  still  fluid.  Once  asolid crust has formed we may assume that at least the surface field is  frozen, with further evolution taking place only on amuch longer diffusive time-scale. It is  thought that the  crustwill  not begin to  form until around100s  after  the  collapse. \citep{spruit06}.

Once the proto-NS has settled down into a quasi-stable state some time after birth, the evolution of the crustal magnetic field is mainly driven through Hall drift \citep{jones88,gourg15}, and later $(\gtrsim 10^{5} \text{ yrs})$ by Ohmic dissipation \citep{gold92,cumm04}. Hall drift is also expected to drive cascades in NS crusts, which in turn can further generate strong, small-scale magnetic structures (`spots') \citep{rhein02,gep03,kojima12,marchant14,li16,suv16}. Recently, \cite{gourg16} showed that, for initial conditions corresponding to a turbulent, crustal magnetic field in a young magnetar with characteristic strength $B \sim 5 \times 10^{14} \text{ G}$, the magnetic energy can decay via Hall and Ohmic evolution by a factor $\sim 2$ within a few $\sim \text{ kyr}$, significantly redistributing energy among high-order multipoles $(10 \lesssim \ell, m \lesssim 40)$ in the process.

As such, even on shorter timescales of $\lesssim 10^{2} \text{ yr}$, the field evolves non-trivially (especially if the natal field $B \gtrsim 10^{15} \text{ G}$ since the Hall time $\tH \propto B^{-1}$), all the while facilitating the growth of zones of magnetic stress throughout the crust \citep{thomp02,lan19}. If the stress within a certain patch exceeds a critical threshold, the crust may locally cease to respond elastically and potentially crack \citep{chug10,lander15,chug18}. Such events might lead to sequences of localised crustquakes, expelling energy at seemingly uncorrelated intervals \citep{lander16,thomp17}, thereby bypassing the intrinsic vs. extrinsic concerns of \cite{li19}.

% For a highly multipolar and strong magnetic field, wherein the crust may experience excessive stresses at many locations corresponding to local maxima of the Lorentz force, many such crustquakes can occur. It is possible that FRBs could be triggered by these small fractures.

% which may be linked to bursting activity \citep{rea11}.

% Indeed, phase resolved spectroscopy reveals that some magnetars have small-scale magnetic field whose strength eceeds their large scale 

This paper is organised as follows. In Section 2 we introduce some phenomenological aspects of repeating FRBs and discuss their possible connection to crustquakes in young magnetars. In Section 3, a discussion of Hall drift and its role in generating magnetic stresses is given, followed by a brief introduction to the von Mises theory of elastic failures in a crust. A simple model is then presented in Section 4, illustrating how one may relate magnetic reconfigurations to quake geometries and energetics. Some discussion is offered in Section 5.

\section{Repeating fast radio bursts}

To date, there are two sources which are known to emit \emph{repeated} FRB signals, namely FRB 121102 \citep{spitler14,tend17,scholz17,gajjar18} and FRB 180814.J0422+73 \citep{chime19,wang19,yang19}. While the second of these has only been recently discovered, which means that statistical analyses are limited, $170$ bursts from the former object have been recorded, from which statistically significant conclusions about the waiting times (Sec. 2.1) and the burst energetics (Sec. 2.2) can be drawn \citep{opp18,li19}.

\subsection{Waiting times}

From data analysis of FRB 121102, \cite{li19} found little to no correlation between burst waiting times, energetics, or duration. In particular, the lack of a correlation between burst energetics and the waiting times between successive bursts is suggestive that the trigger mechanism is either extrinsic [e.g. a neutron star embedded within an asteroid field \citep{dai16}], or that small scale intrinsic (i.e. non global) mechanisms are responsible. The source also appears to be episodic, alternating between epochs of activity, wherein many bursts occur [e.g. $93$ bursts in FRB 121102 were found within 5 hours of observing time by \cite{zhang18b}] and quiescence, wherein no detectable bursts are released for several hours or more \citep{opp18,price18}. 

%%%MENTION THAT 10^-2 IS ONLY A FEW AND THE LATER ONE IS MUCH MORE.
\cite{li19} further found that the waiting times appear to be bimodal, clustering around $10^{-2} - 10^{-3}$ s and (more so) at $\sim 10^{2}$ s. The former of these is of the order of the Alfv{\'e}n crossing time,
\begin{equation} \label{eq:alfv}
t_{\text{A}} \sim 10^{-3} \left( \frac {\rho} {10^{13} \text{ g cm}^{-3}} \right)^{1/2} \left( \frac {L} {10^{5} \text{ cm}} \right) \left( \frac {B} {10^{15} \text{ G}} \right)^{-1} \text{ s},
\end{equation}
for characteristic crustal density $\rho$ and length-scale $L \sim R_{\star} - R_{c} \approx 0.1 R_{\star}$, with $R_{c}$ and $R_{\star}$ denoting the crustal and stellar radii, respectively. It is well known that a variety of magnetohydrodynamic instabilities can occur over a few $t_{\text{A}}$ in some systems \citep{kokk14}. Rapid rotation can, however, delay the onset of magnetic instabilities by a factor $\sim P^{-1}$ for spin period $P$ \citep{spruit06}, so that the instability time-scale associated with \eqref{eq:alfv} could easily be of the order $\gtrsim 10^{-2}$ s in lighter sections of crust $(\rho \lesssim 10^{11} \text{ g cm}^{-3})$ for $P \gtrsim 10 \text{ ms}$. Furthermore, it has been shown that an overstability of global elastic modes may by caused by the collective shearing of locally overstressed patches, releasing energy over time-scales $\sim 10^{2}$ s \citep{thomp17}. This latter timescale may be related to the longer mode of the waiting time distribution. In any case, the magnetic field is likely to play a prominent role.

%In any case, the similarity between the Alfv{\'e}n time \eqref{eq:alfv} and the clustered burst waiting time suggests that, at least for the intrinsic mechanism scenario, that the magnetic field is likely to play a prominent role.

%Putting these pieces of evidence together an
%or even up to the longer timescale in deeper regions with weaker fields.

%In any case, %Furthermore, for the reasons discussed in the introduction, the field is likely to be highly multipolar, which is consistent

%In the outermost layers of the crust, the crust melts and cannot provide shear stress.

%The Hall time-scale cannot be made arbitrarily small; however,because at very low electron densities shear stresses in the crustare no longer able to balance the Lorentz forces.

\subsection{Energetics}

While brighter bursts might be intensified by plasma lensing \citep{cordes17}, if we assume isotropic emission, then the observed fluence $F$ is related to the FRB energy $\EF$ through \citep{zhangs18}
\begin{equation} \label{eq:frbenergy}
\left( \frac {\EF} {1.2 \times 10^{39} \text{ erg}} \right) \approx \left( \frac {F} {\text{Jy ms}}  \right) \left( \frac {\nu} {\text{GHz}} \right) \left( \frac {D} {\text{Gpc} } \right)^{2} \left( 1 + z \right)^{-1},
\end{equation}
where $\nu$ is the source frequency, $D$ is the luminosity distance, with redshift $z \approx 0.2$ $(D \approx 1 \text{ Gpc})$ for FRB 121102 \citep{chatt17} and $z \lesssim 0.11$ ($D \lesssim 0.5 \text{ Gpc}$) for FRB 180814.J0422+73 \citep{yang19}. Any mechanism that is proposed to instigate FRB behaviour should predict energetics which are consistent with \eqref{eq:frbenergy}.

\cite{wang18} found that if one fits a power-law to the number distribution of burst energies $N(E)$ for FRB 121102, $N(E) \propto E^{-\gamma}$, then the best fit is obtained with the value $\gamma = 2.16 \pm 0.24$ at the $95\%$ confidence level (see their Fig. 1); see also \cite{lu16} who find $1.5 \lesssim \gamma \lesssim 2.2$ when including data from other FRBs and \cite{wangxx}. \cite{wang18} drew a comparison with the statistics for earthquakes, which are well described by the \cite{gutenberg56} law $N(E) \propto E^{-2}$. Though far from conclusive, the waiting time distribution, which was found to be consistent with a Poisson or Gaussian distribution [though cf. \cite{opp18}], also agrees with empirical relationships for seismicity rates of earthquakes \citep{utsu95}. 

\subsection{Radio emission}

{How might a quake in a neutron star source coherent radio emission? As an example, \cite{wad19} proposed that the radio emission might originate from the closed field line zone within the magnetosphere surrounding a magnetar. If the magnetospheric twist is sufficiently low, so that the plasma above the active region is unable to screen the accelerating electric field, generated by a crust yielding event, the plasma density in the closed field zone would be low enough to permit the escape of $\sim$ GHz radio waves. This is similar to the classical picture of coherent emission from pulsars [see also \cite{blaes89}]. Other mechanisms have been proposed [see e.g. \cite{belo17}], which are also consistent with a quake picture \citep{petroff19}. }

%
% warrant a further comparison between these two phenomena, which we aim to explore through a simple model (Sec. 3).

\section{Crustquake model}

%over the course of a few Alfv{\'e}n crossing times \eqref{eq:alfv} or later through collective shear overstabilities \citep{thomp17},

Putting the pieces of evidence surrounding timescales and energetics together, it is possible that tectonic activity, resulting in the expulsion of energy from the crust of a young magnetar, may be connected to repeating FRBs \citep{wang18}. Furthermore, if a topologically complicated, quasi-equilibrium magnetic structure is frozen into the crust $\sim 10^{2} \text{ s}$ after birth, complicated fracture geometries are likely to result as the field then relaxes to a more stable state over the longer dynamical and diffusion timescales.

To investigate this possibility further it is necessary to understand which mechanisms, if any, may be responsible for driving both local and global magnetic field evolution over the observed age $\lesssim 10^{2}$ yr of FRB 121102 (Sec. 3.1), and to further detail a model of the crustal failures that might result (Sec. 3.2).

\subsection{Hall drift}

Hall drift, namely the process of field line advection due to the generation of an electric current from magnetic flux transport by mobile electrons, is believed to play a crucial role in NS crustal field evolution during the first $\lesssim 10^{5}$ yr after formation \citep{rhein02,gep03,kojima12}, after which Ohmic dissipation takes over \citep{gourg16} [though cf. \cite{cumm04}]. As initially put forth by \cite{gold92} [and later confirmed through numerical simulations \citep{rhein02,gep03,kojima12,marchant14,li16}], turbulent cascades resulting from the Hall drift can transfer energy from larger to smaller scales within the crust. This suggests that a star born with a multipolar field may have it persist, at least over the Hall timescale (see below). Furthermore, as the field relaxes, energy redistributions occur not only between the multipolar components, but also between the poloidal and toroidal components \citep{vig12,gep14}, which can further instigate the formation of anisotropic, overstressed zones.

In general, the Hall timescale reads \citep{gourg15}
\begin{equation} \label{eq:hall}
\tH \sim 640 \left( \frac{n_{e}}  {10^{34} \text{ cm}^{-3}} \right) \left(  \frac{L}  {10^{5} \text{ cm}} \right)^{2} \left( \frac{B}  {10^{15} \text{ G}} \right)^{-1} \text{ yr},
\end{equation}
where $n_{e}$ is the electron number density, which varies between $10^{34}  \lesssim n_{e} \lesssim 10^{36}$ throughout the crust $R_{c} \leq r \leq R_{\star}$. Clearly, over short length-scales $(L \lesssim 10^{5} \text{ cm})$ or for locally intense magnetic fields $(B \gtrsim 10^{15} \text{ G}$), the Hall time \eqref{eq:hall} could be of the order of a century or less \citep{cumm04}, which coincides with the predicted age for the object associated with FRB 121102 \citep{bower17,metzger17}. 

In line with the theory of \cite{gold92}, \cite{gourg16} found that strongly multipolar fields $(10 \lesssim \ell, m \lesssim 40)$ both persist and are generated over a few $\sim \text{ kyr}$, i.e. over several Hall times \eqref{eq:hall}, in the crusts of strongly magnetised $(B \gtrsim 5 \times 10^{14} \text{ G})$ NSs [see also \cite{dall09}], and further that substantial energy redistributions, between the individual multipolar components, occurs during this time. In regions of concentrated field (e.g. toroidal sections), redistributions occur even more rapidly as $\tH \propto B^{-1}$ \citep{vig12,gep14}.

%Note that, while we have only discussed Hall drift, other mechanisms may also be responsible for significant field rearrangements in the crusts of young magnetars during the first $\lesssim 10^{2}$ yr of their lives, such as through thermomagnetic instabilities which convert thermal energy into magnetic energy \citep{urpin86,gepp17}, or ambipolar diffusion resulting from direct Urca processes \citep{pass17}.

\subsection{Fracturing crusts}

%Suppose that we assume that a cracking crust is indeed (at least partially) responsible for FRB behaviour. In light of the mechanisms discussed in the introduction, we further assume that a rearranging magnetic field is the main mechanism which facilitates the tectonic activity. How can one determine where and when a crust fracture occurs? 
The problem of determining the yieldability of an elastic medium was considered by \cite{vonm13} over a century ago. Modelling the NS crust as an elastic medium, the von Mises criterion may therefore be applied to study its ability to support stresses, as has been done by several authors, e.g. \cite{pons11,pons112,johnson13,lander15}, and \cite{lan19} [though cf. \cite{chug18}]. The von Mises criterion for the critical stress beyond which an isotropic crust no longer responds elastically, and may therefore crack, reads \citep{lander15}
\begin{equation} \label{eq:vonmises}
\sqrt{ \tfrac {1} {2} \sigma_{ij} \sigma^{ij} } \gtrsim \sm,
\end{equation}
where $\boldsymbol{\sigma}$ is the elastic strain tensor and $\sm$ is the maximum breaking strain of the crust. In general, the ions in the crustal layers interact via Coulomb potentials which are screened by the mobile, degenerate electrons, and form a crystal, the particulars of which determine the elastic properties of the crust \citep{horo09,chug10,hoff12,chug18}. Although depending on various factors, such as the degree of hydrostatic pressure anisotropies, molecular-dynamics simulations performed in the aforementioned studies find $10^{-3} \lesssim \sm \lesssim 10^{-1}$. 

While the strain tensor $\boldsymbol{\sigma}$ is naturally a dynamic quantity, in the sense that it depends on the fluid motions, we can consider a quasi-static evolution in the magnetic field to determine which sections of crust might have yielded due to magnetic stresses over some period of time \citep{lander15}. In particular, we consider pairs of magnetic field configurations, each of which consists of an `initial' field $\boldsymbol{B}_{i}$ and a `final' field $\boldsymbol{B}_{f}$. In other words, we assume that the `initial' field has decayed into some new configuration via Hall drift or otherwise, which we call the `final' field, and explore the resulting crustal stress. %A specific decay mechanism, Hall drift, is explored in Sec. 3.2.

%10^{-2} \lesssim t
Considering magnetic stresses alone {(see below)}, the von Mises criterion \eqref{eq:vonmises} reads\footnote{
{The surface temperature of a young neutron star can evolve rapidly due to plasmon decay during the first $ \lesssim 10$ yrs after birth, and subsequently over the next $\lesssim 10^{3}$ yrs due to a combination of electron-nucleus and neutron-neutron bremsstrahlung \citep{gned00}. These thermal relaxation processes can alter the crustal structure, allowing expansions, contractions, and equation of state shifts to occur before and during the Hall time \eqref{eq:hall}. It is for this reason we have used the form \eqref{eq:vonb} for the magnetic stress from \cite{lander15} rather than the form proposed in \cite{lan19}, whose prefactors differ by a factor $\sim 2$ by not allowing expansions/contractions to source $\sigma_{ij}$. In any case, these factor $\sim 2$ differences do not qualitatively affect our conclusions (see Sec. 4.4).}}

\begin{figure}
%\centering
\includegraphics[width=0.473\textwidth]{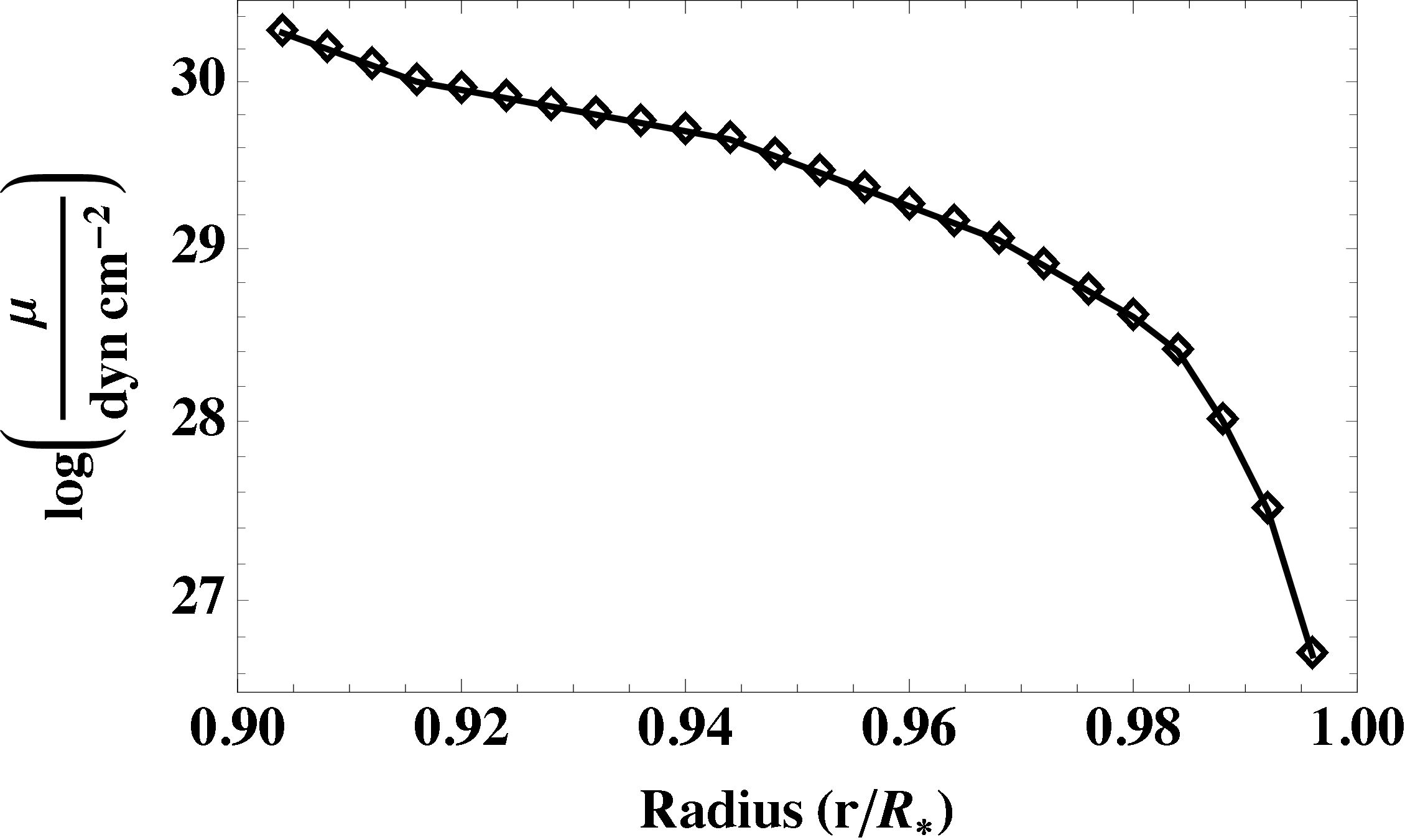}
\caption{Shear modulus $\mu$ in a NS crust ($0.9 \leq r / R_{\star} \leq 1$) used in this paper. The profile was determined using the method described in Lander et al. (2015) from the data provided in Douchin \& Haensel (2001), Horowitz \& Hughto (2008), and Kaminker et al. (2009). The profile is similiar to others adopted in the literature, e.g. Sotani et al. (2007). \label{shearmodulus}
}
\end{figure}

%We have chosen to use the form \eqref{eq:vonb} for the von Mises stress in \cite{lander15} rather than the result of \cite{lan19}, which differ by a factor $\sim 2$ in their prefactors. The former authors implicitly allow yielding to be partially sourced by crustal contraction or expansion, which may occur in young n
%Note that the prefactors of the terms within \eqref{eq:vonb} differ between the works of \cite{lander15} and \cite{lan15}, where the former implicitly assume 

\begin{equation} \label{eq:vonb}
\begin{aligned}
\sqrt{ \tfrac {1} {2} \sigma_{ij} \sigma^{ij} } &= \frac {\sqrt{ |\bb_{f}|^2 |\bb_{i}|^2 + \tfrac {3} {2} |\bb_{f}|^4 + \tfrac {3} {2} |\bb_{i}|^{4} - 4 \left( \boldsymbol{B}_{f} \cdot \boldsymbol{B}_{i} \right)^{2} }} {8 \pi \mu} \\
&\gtrsim \sm,
\end{aligned}
\end{equation}
where $\mu$ is the shear modulus. Following \cite{lander15}, we estimate the shear modulus throughout the crust ($R_{c} \leq r \leq R_{\star}$) from the molecular-dynamics simulation data provided by \cite{horo08} supplemented by the liquid drop equation of state of \cite{douch00}. The temperature profile of \cite{kamin09}, who modelled cooling via neutrino emission and thermal conduction, is further used to estimate the Coulomb coupling parameter. The fit we obtain, identical to that of \cite{lander15}, and in quantitative agreement with others [e.g. the profile used by \cite{sotani07}], is shown in Fig. \ref{shearmodulus}. %NEED TO THINK ABOUT THIS; THE CURVE IS FOR AGE t=1KYR, DOES THIS MATTER?

{It is important to note that other, fluid-dynamical mechanisms may contribute to the overall crustal strain, particularly the spin-down \citep{baym71}. Using coupled crust-core models and elastic deformation theory, \cite{cut03} have estimated the crustal strain due to neutron star precession and spin-down, which, for a newborn, millisecond magnetar, reads [see their equation (73) in particular]}
\begin{equation} \label{eq:precession}
\left( \tfrac{1}{2} \sigma_{ij} \sigma^{ij} \right)^{1/2}_{\text{sd}} \approx 5 \times 10^{-2} \left( \frac {P} {\text{ ms }} \right) \left( \frac{ P_{\text{p}}} {10^{2} \text{ s}} \right)^{-1} ,
\end{equation}
{where $P_{\text{p}}$ is the precession frequency, given by $P_{\text{p}} \approx P/ \epsilon$ \citep{gog15}, where $\epsilon \lesssim 10^{-5}$ (see Sec. 5) is the oblateness parameter. In general, we find that the spin-down strain estimated in \eqref{eq:precession} is sub-leading compared to the magnetic deformation unless $\epsilon \gtrsim 10^{-4}$, which does not occur unless the star is significantly oblate, requiring a strong poloidal field with $B \gtrsim 10^{16} \text{ G}$ \citep{gual10,lasky13}. In any case, we find that the stars are primarily prolate for our models (see Table \ref{tab:ellipdata} below), and that magnetic stresses dominate over spin-down stresses \eqref{eq:precession} (see Sec. 4.4), so we ignore the latter contribution.}

%day) model with crustal strains sourced by the magnetic field, and varying the other, ‘before’ (original) configura- tion – i.e. the initial star with its relaxed crust. We assume the ‘before’ field has decayed into the ‘after’ field – so that the greater the difference in magnetic energy between these models, the larger the region of the crust that should be strained to the point of yielding. We also explore the effect of varying the breaking strain and the ‘after’ field strength

\cite{lander15} additionally found that the energy released in a shallow $(d \lesssim 0.5 R_{c})$ quake [i.e. the magnetic energy stored in regions wherein \eqref{eq:vonmises} is satisfied] can be well approximated by
\begin{equation} \label{eq:equake}
\left( \frac {\Eq} {4 \times 10^{39} \text{ erg}} \right) \approx \left( \frac {\sm} {10^{-2}} \right) \left( \frac {d} {10^{-1} R_{c}} \right)^{2} \left( \frac {l} {10^{3} \text{ cm} } \right),
\end{equation}
where $d$ is the quake penetration depth, and $l$ is the fracture length. Clearly, in order to match the energy requirements $\Eq \gtrsim \EF$, only slight fractures of the crust are necessary $(l \lesssim 10^{-3} R_{\star}, d \ll R_{c})$.

In general, the geometry of a crustal failure depends on the magnetic field topology, as is evident from \eqref{eq:vonb} \citep{link00}. Localised, small fractures of the crust are therefore possible if the stress \eqref{eq:vonb} is strongly anisotropic, as would occur for an evolving, multipolar magnetic field. Bursts with characteristic energy \eqref{eq:equake} will then be released as the spatially dependent criterion \eqref{eq:vonmises} is met in certain, disconnected sections of the crust. %e.g. over the Hall time \eqref{eq:hall}.

%\subsection{Rarity}

%Thompson C., Duncan R.C., 1995, ApJ 275, 255 [maybe cite this too for \Eq estimate...%
%Basically, we found that the behaviors of the repeat- ing FRB 121102 are earthquake-like. The distribution of burst energy exhibits a Gutenberg-Richter power law form which is a well-known earthquake distribution. And the distribution of waiting time, can be characterized as a Poissonian or Gaussian distribution, which are con- sistent with earthquakes as well as the local correlated aftershock sequence. The possible origins of the repeater are discussed including crustal activity of a magnetar and solidification-induced stress of a new-born SS. Both possible origins might be associated with SGRs which are difficult to detect at cosmological distance. Statis- tic distributions of burst energy and duration time show that FRB 121102 is very similar to SGR 1806 − 20 (Wang & Yu 2017). Also, SGR 1806 − 20 share some dis- tinctive properties with earthquake that indicates SGRs are indeed powered by starquakes (Cheng et al. 1996), and the giant flares of SGRs are suggested to be quake- induced (Xu et al. 2006).

\section{Demonstrative models}

In this section, we present some simple models for initial, $\bb_{i}$, and final, $\bb_{f}$, field states which aim to capture the major phenemonological features observed in the numerical simulations of \cite{gourg16}. Through this, we can examine the geometry of fractures, and thus further assess the viability of the crustquake explanation for repeating FRBs initially considered by \cite{wang18}.

\subsection{Magnetar crustal field}

Although we take inspiration for our field configurations through the simulations of \cite{gourg16}, who found that non-axismmyetry played an important role for the evolution, since we only consider a quasi-static sequence (i.e. a $\bb_{i}$ and a $\bb_{f}$) of magnetic fields here, it is reasonable to consider fields which have energies similar to those in \cite{gourg16}, but which are axisymmetric. In spherical coordinates ($r,\theta,\phi$), an axisymmetric vector field can always be split into a mix of poloidal and toroidal components, viz. 
\begin{equation} \label{eq:bfield}
\bb = \tilde{B} \left[ \nabla \alpha \times \nabla \phi + \left(\frac {E^{p}} {E^{t}} \frac {1 - \Lambda} {\Lambda} \right)^{1/2} \beta(\alpha) \nabla \phi \right],
\end{equation}
where $\tilde{B}$ is the characteristic field strength, $\alpha = \alpha(r,\theta)$ is a scalar flux function, and $\beta$, which describes the spatial variation of the toroidal component, is a function of $\alpha$ only \citep{c56,mlm13,msm15}. In expression \eqref{eq:bfield}, we have introduced the constructions
\begin{equation} \label{eq:epol}
E^{p} = \frac {1} {8 \pi} \int_{V} dV \left[ \left( \frac {1} {r^2 \sin\theta} \frac {\partial \alpha} {\partial \theta} \right)^2 + \left( \frac {1} {r \sin\theta} \frac {\partial \alpha} {\partial r} \right)^2 \right],
\end{equation}
and
\begin{equation} \label{eq:etor}
E^{t} = \frac {1} {8 \pi} \int_{V} dV  \frac {\beta(\alpha)^2} {r^2 \sin^2\theta},
\end{equation}
which represent the poloidal and toroidal field energies stored within \emph{crustal} volume $(R_{c} \leq r \leq R_{\star})$ $V$, respectively. In \eqref{eq:bfield}, the parameter $0 < \Lambda  \leq 1$ characterises the relative strength between the poloidal and toroidal components, e.g. $\Lambda = 0.5$ gives a field which has an equal poloidal-to-toroidal field strength ratio: $E^{p} = E^{t}$.

The stream function $\alpha$ can be decomposed into a sum of multipoles \citep{mlm13},
\begin{equation} \label{eq:alphal}
\alpha = \sum_{\ell} \alpha_{\ell} = \sum_{\ell} \kappa_{\ell} f_{\ell}(r) Y_{\ell 0}'(\theta) \sin \theta, 
\end{equation}
where the $Y_{\ell m}$ are the spherical harmonics, and the $\kappa_{\ell}$ set the relative strengths between the multipolar components of $\bb$ (see Sec. 4.2). The radial functions $f_{\ell}$ are chosen as polynomials such that $\bb$ from \eqref{eq:bfield} is everywhere finite, continuous both inside the crust ($R_{c} \leq r  \leq R_{\star}$) and with respect to a current-free external field ($r \geq R_{\star}$), and to ensure that the magnetic current $\boldsymbol{J} \propto \nabla \times \bb$ vanishes at the surface $(r=R_{\star})$ \citep{mmra11,mlm13,msm15}.  

We further choose

\begin{equation} \label{eq:betafn}
 \beta(\alpha) =
\begin{cases}
(\alpha - \alpha_{c})^2 / R_{\star}^3&\textrm{for }\alpha\geq \alpha_{c},\\
0&\textrm{for }\alpha < \alpha_{c},
\end{cases}
\end{equation}
where $\alpha_{c}$ is the value of $\alpha$ that defines the last poloidal field line that closes inside the star. The simple quadratic \eqref{eq:betafn} for $\beta$ ensures that the toroidal field is confined within an equatorial torus [in line with the numerical simulations of \cite{braith06,b09}], bounded by the last closed poloidal field line defined by $\alpha_{c}$ \citep{mlm13,msm15,suv16,suv18}.

\subsection{Energy distribution}

In this subsection we present explicit models for the generic fields \eqref{eq:bfield} defined in the previous section. In particular, we wish to consider crustal fields in young magnetars which are initially `turbulent' as discussed in the Introduction, i.e. fields that begin as a superposition of high order magnetic multipoles with a roughly uniform energy distribution across $\ell$, which then evolve toward a more traditional dipole-dominated configuration \citep{gourg16}. This assumption is further supported by observations of millisecond magnetar braking indices $n$ finding $n \lesssim 3$ \citep{lasky17}, suggesting that the dipole moment of young magnetars increases with time \citep{gourg15}. 

In general, we relate the energy between each of the multipolar components in \eqref{eq:alphal} through the general expression
\begin{equation} \label{eq:energydist}
E^{p}_{\ell,\xi} = \frac {\ell^{-\xi}} {H_{\ell_{\text{max}},\xi}} E_{0},
\end{equation}
where $1 \leq \ell \leq \ell_{\text{max}}$, $H_{\ell_{\text{max}},\xi} =\sum_{i=1}^{\ell_{\text{max}}} i^{-\xi}$ are the generalised harmonic numbers, $E^{p}_{\ell,\xi}$ is the energy stored within the $\ell$-th poloidal component, and $E_{0}$ is a canonical energy `budget' for the magnetic field, which we set as $E_{0} = 2.2 \times 10^{46} \text{ erg}$. Integrating the components of the field \eqref{eq:epol} with \eqref{eq:alphal} and imposing \eqref{eq:energydist} yields a linear system for the $\kappa_{\ell}$. The parameter $\xi$, appearing within \eqref{eq:energydist}, provides a convenient way to assign an energy distribution amongst the multipolar components in the following sense. If we set $\xi = 0$ within \eqref{eq:energydist}, then the multipolar energy distribution is uniform; $E^{p}_{i} = E^{p}_{j}$ for all $1 \leq i,j \leq \ell_{\text{max}}$. The harmonic numbers $H_{\ell_{\text{max}},\xi}$ are normalisation constants, included to ensure that the total poloidal energy is constant between configurations, i.e. $\sum_{\ell} E^{p}_{\ell,\xi} = E_{0}$. In general, each successive multipole has, with respect to its predecessor, an energy ratio
\begin{equation} \label{eq:ratio}
\frac {E^{p}_{\ell+1,\xi}} {E^{p}_{\ell,\xi}} = \frac {\ell^{\xi}} {\left( \ell + 1 \right)^{\xi}},
\end{equation}
which is strictly decreasing for fixed $\ell$ and $\xi > 0$, and strictly increasing for $\xi < 0$. For example, $\xi = 1$ gives $E^{p}_{1} = 2 E^{p}_{2} = 3 E^{p}_{3} = \cdots = \ell_{\text{max}} E^{p}_{\ell_{\text{max}}}$, so that the dipole component has twice as much energy as the quadrupole, and so on. Such a field would be highly-ordered and resembles more of a traditional dipole dominated field, while a field with $\xi \approx 0$ is more `turbulent' in the sense of \cite{gourg16} (see their Fig. S5 in particular). In this way, $\xi$ provides a measure for the extent of field reordering between subsequent configurations. Similarly, the value of $\Lambda$ between successive configurations adjusts as energy is exchanged between the poloidal and toroidal components, which also occurs in Hall drift \citep{vig12,gep14,suv16}. 

\subsection{Models}

We consider three pairs of magnetic fields, each of which consist of an `initial' and a `final' configuration (denoted with subscripts $i$ and $f$, respectively). We assign each field a characteristic strength $\tilde{B} = 10^{15} \text{ G}$ and resolve up to multipoles of order $\ell_{\text{max}} = 27$. The initial and final states are related by an adjustment of their respective $\xi$ and $\Lambda$ values through the assumption that the multipolar components tend to become more ordered on the Hall time \eqref{eq:hall}, as found in \cite{gep14,gourg16}, in the sense that we set $\xi_{f} > \xi_{i}$ within \eqref{eq:energydist}. We further take $\Lambda_{i} < \Lambda_{f}$, so that the final state has less total energy than the initial state (see Sec. 4.4). The properties of the three models (named A, B, and C) used in this paper, in terms of the above parameters, are listed in Table \ref{tab:simdata}.

\begin{table*}
\caption{Properties of the magnetic field configurations used within this paper. The fields, each of which have a poloidal-to-toroidal field energy ratio $\Lambda$, are defined through the Chandrasekhar (1956) decomposition \eqref{eq:bfield}. The multipolar components have their energies distributed according to relationship \eqref{eq:energydist} with $E_{0} = 2.2 \times 10^{46} \text{ erg}$. All fields have the same characteristic strength $\tilde{B} = 10^{15} \text{ G}$ and maximum multipolar resolution $\ell_{\text{max}} = 27$.}
  \begin{tabular}{llcccc}
  \hline
Model  & $\xi_{i}$ & $\xi_{f}$ & $\Lambda_{i}$ & $\Lambda_{f}$ \\
\hline
A & $0.2$ & $0.4$ & $0.4$ & $0.6$ \\
B & $-0.5$ & $-0.1$ & $0.45$ & $0.55$ \\
C & $0.7$ & $1.0$ & $0.6$ & $0.7$ \\
\hline
\end{tabular}
\label{tab:simdata}
\end{table*}

%HAVE SOME TABLES WITH PARAMETER VALUES ETC. PLUS SOME PLOTS OF THE BEFORE AND AFTER FIELDS SIDE BY SIDE.

%a superposition of odd and evenmultipoles is symmetric about the magnetic axis, but not symmetric about the equator. he pure multipoles are north-south symmetric or antisymmetric, so a superposition of even (odd)multipoles is always symmetric (antisymmetric)The weakening of the other multipoles means that the degree is north-south hemisphere asymmetry is reduced. (i.e. meant to emulate a `turbulent' field inside a proto-magnetar, as described in the introduction),in line with the age of the hypothetical NS within FRB 121102 \citep{bower17,metzger17}. 

\begin{figure*}
%\centering
\includegraphics[width=\textwidth]{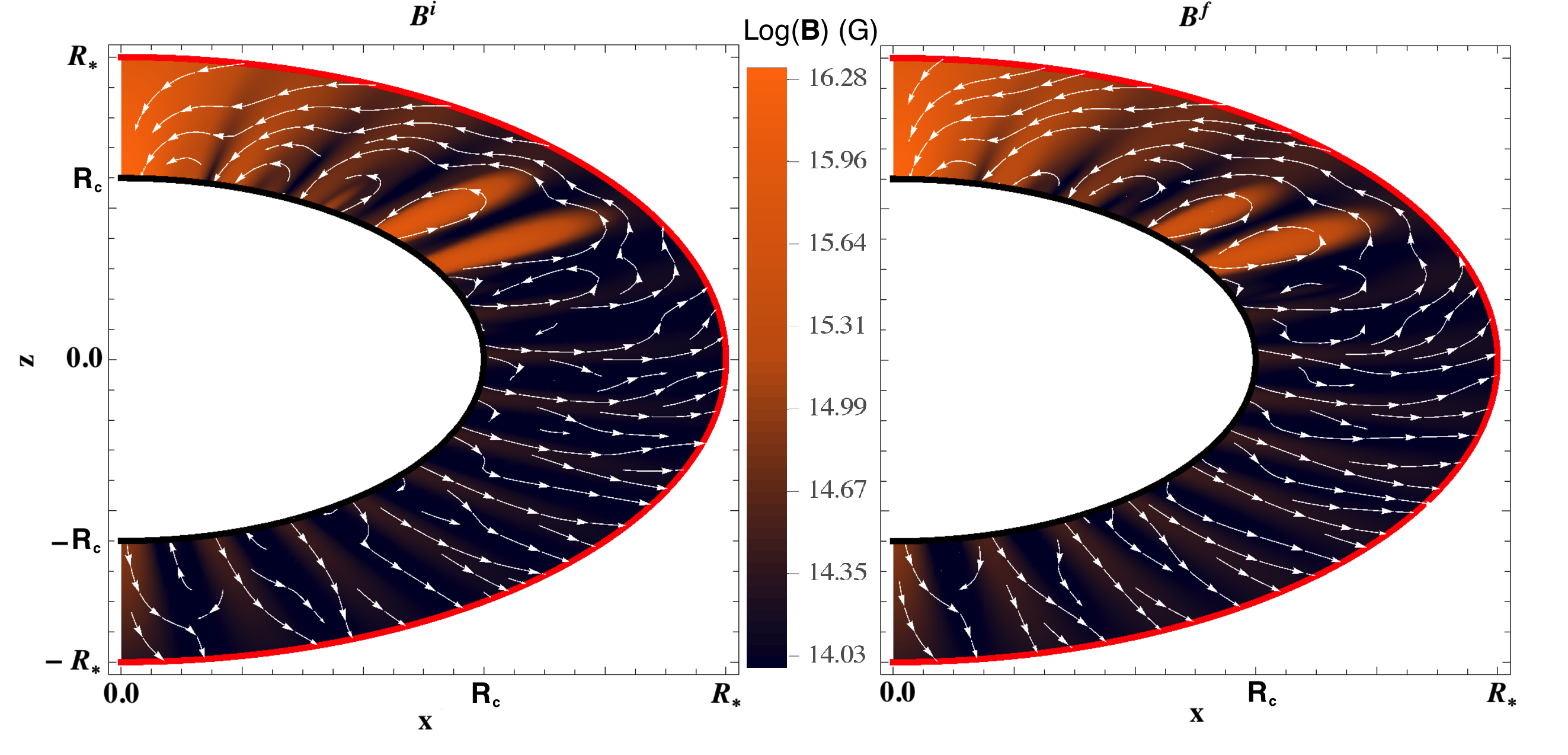}
\caption{Initial (left panel) and final (right panel) field configurations associated with model A. The colour scale shows the magnitude of the magnetic field  $|\bb|$, with brighter shades indicating a stronger field. The stellar surface $(r = R_{\star})$ is shown by the red curve, while the crust-core boundary $(r = R_{c} \equiv 0.9 R_{\star})$ is shown by the black curve. The crust has been stretched by a factor $4$ for improved visibility. (Any apparent field line discontinuites are plotting artifacts.) \label{maga}
}
\end{figure*}

In Figure \ref{maga} we present the `initial' (left panel) and `final' (right panel) configurations corresponding to model A. For this case, the energies stored in the multipolar components of the initial magnetic field are roughly uniform with $\ell$, though the dipole component is strongest, containing $6.0\%$ of the total energy as $\xi_{i} = 0.2$. For this model, the Hall time \eqref{eq:hall} reads $\tH \lesssim 60 \text{ yr}$ in the northern hemisphere, while $\tH \gtrsim 10^{3} \text{ yr}$ in the southern hemisphere (see below). The magnetic structure is also complicated, exhibiting the plume-like features characteristic to multipolar fields \citep{jackson62}. During the Hall time, we assume that the field has evolved from this disordered configuration towards a more stable configuration in which the dipole and other, lower $\ell$ components grow at the expense of the high-$\ell$ multipoles. We set $\xi_{f} = 0.4$ so that the dipole component of the final state contains $9.1\%$ of the total energy. We see that, for both configurations, the field is strongest at the north pole $(|\bb| \approx 2 \times 10^{16} \text { G})$, and in the respective toroidal regions $(|\bb| \approx 10^{16} \text { G})$ just above the equator. Though the initial and final fields are qualitatively similar, even small reconfigurations in regions with strong $|\bb|$ can exert significant stresses on the crust (see Sec. 4.4). 

In general, even order multipoles are symmetric about the equator, while odd order multipoles are anti-symmetric. This implies that a field which is a general superposition of odd and even order multipoles will be much stronger in the northern hemisphere. As such, as the field becomes more ordered (increasing $\xi$), $|\bb|$ tends to increase in the southern hemisphere \citep{mlm13,msm15}. Furthermore, in each model, the final toroidal field is relatively weaker as $\Lambda_{i} < \Lambda_{f}$, and we have that the overall energy has decreased by $\approx 10^{45} \text{ erg}$ [a fraction of which may be released through crustquakes; see Sec. 4.4], which can further facilitate a reordering amongst the poloidal terms \citep{vig12,gep14,suv16}. 

\begin{figure*}
%\centering
\includegraphics[width=\textwidth]{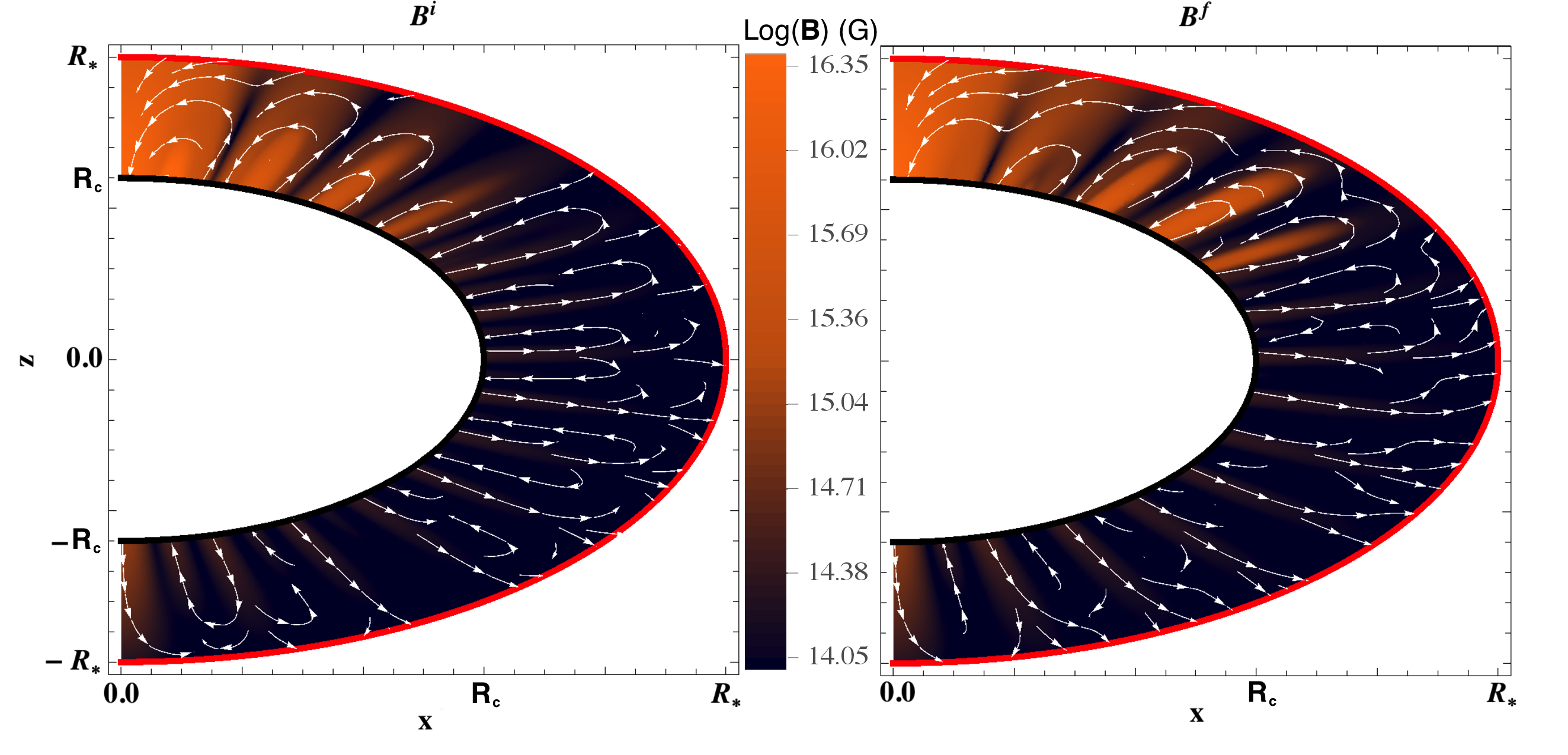}
\caption{As in Fig. \ref{maga}, though for the magnetic fields defined within model B. \label{magb}
}
\end{figure*}

\begin{figure*}
%\centering
\includegraphics[width=\textwidth]{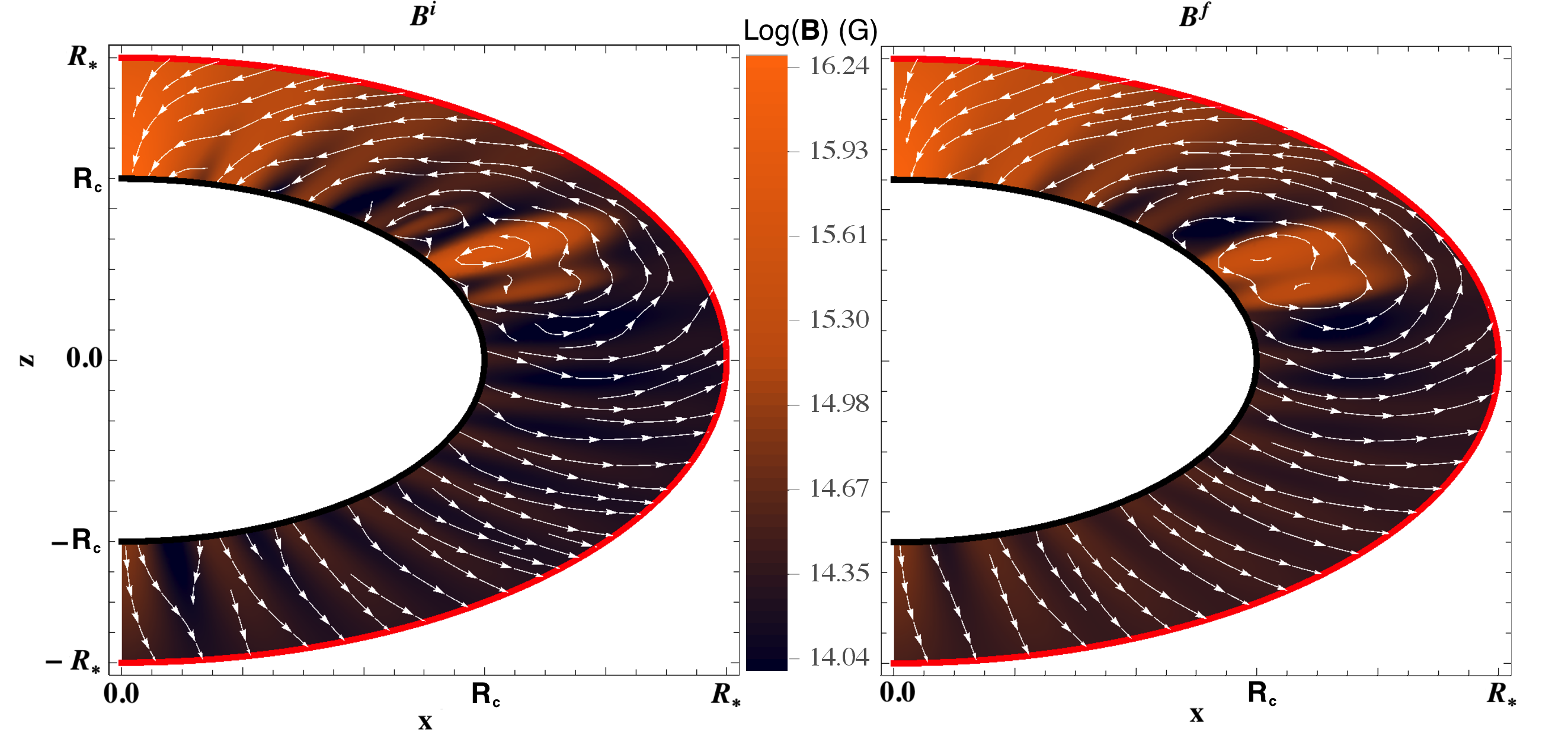}
\caption{As in Fig. \ref{maga}, though for the magnetic fields defined within model C. \label{magc}
}
\end{figure*}

%Figures \ref{magb} and \ref{magc} are similar to Fig. \ref{maga}, except that we show the magnetic field structures for models B and C, respectively. 
For model B, depicted in Figure \ref{magb}, we have that the higher multipole terms actually dominate over the dipole component $(\xi < 0)$, so that the filamentary structure is even more evident, especially due to the closed magnetic loops which confine the toroidal field. As the field evolves towards one with a stronger dipole field $(\xi_{f} = -0.1 > \xi_{i})$, the toroidal field spreads equatorially, as expected. Figure \ref{magc} (model C) depicts a situation wherein the field is largely dipolar $(E_{\text{dip}} = 0.16 E_{0})$ initially, though then becomes even more dipole dominated, with the dipole field containing $26\%$ of the energy in the final state. For this configuration we see that the higher multipoles are less visible, and the field resembles a dipole field with an off-axis toroidal field [similar to the fields described in \cite{mlm13}], especially for the final state with $\xi_{f} = 1.0$. The growing strength of the field in the southern hemisphere is more evident here, even though the field is relatively weak $(|\bb| \approx 5 \times 10^{14} \text{ G})$ there.

\subsection{Fractures and energetics}

%Depending on the fracture geometry, certain quakes have different characteristic energies according to \eqref{eq:equake}.

\begin{figure}
%\centering
\includegraphics[width=0.473\textwidth]{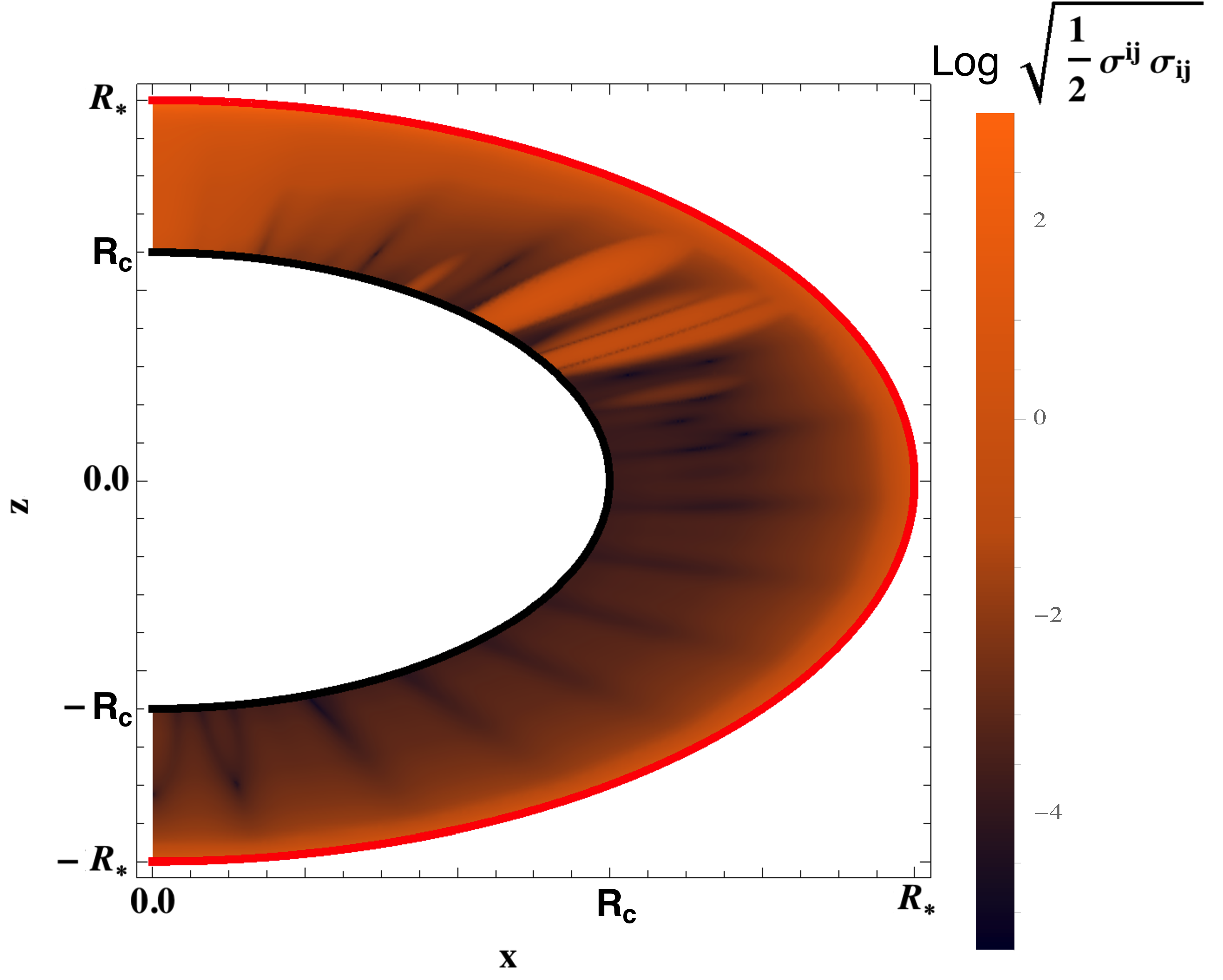}
\caption{Contracted elastic strain tensor, defined within \eqref{eq:vonb}, for model A. The colour scale shows the magnitude, with brighter shades indicating greater stress. The stellar surface $(r = R_{\star})$ is shown by the red curve, while the crust-core boundary $(r = R_{c} \equiv 0.9 R_{\star})$ is shown by the black curve. Regions wherein $\sqrt { \tfrac {1} {2} \sigma_{ij} \sigma^{ij}} \geq \sigma_{\text{max}}$ cease to respond elastically and may crack according to the von Mises criterion \eqref{eq:vonmises}. The crust has been stretched by a factor $4$ for improved visibility. \label{sigmaa}
}
\end{figure}

In Figure \ref{sigmaa} we plot the contracted strain tensor $\tfrac {1} {4} \sqrt{\sigma_{ij} \sigma^{ij}}$ \eqref{eq:vonb} for the magnetic fields given by model A. Even with a conservative value of $\sigma_{\text{max}} = 10^{-1}$, we see that the contracted elastic strain tensor can easily exceed the von Mises yielding threshold in several zones throughout crust \citep{lander15,lan19}. The fracture geometry is tied to both the initial and final field states, especially through their relative multipolar strengths, as can be seen through the  filamentary angular structure in Fig. \ref{maga}. Because of the differing north-south symmetry properties between odd and even order multipoles, crustal fractures are separated by `magnetic walls'. In particular, the strain on the very outer layers across the crust is large, so that many long $(l \approx R_{c})$ but shallow $(d \ll R_{c})$ quakes may occur during the Hall time \eqref{eq:hall}. Similarly, deep $(d \lesssim R_{c})$ but short $(l \approx 10^{3} \text{ cm})$ fractures may develop in between each of these magnetic walls at rates which depend on the local Hall time \eqref{eq:hall}, and therefore implicitly through the initial and final values of $\xi$. Furthermore, the toroidal fields may independently instigate quakes, as the azimuthal magnetic fields, and hence stresses through \eqref{eq:vonmises}, are highly concentrated in these regions. Since the toroidal fields are confined within the closed field lines, a changing poloidal field geometry shifts the toroidal rings, which contributes to fracture growth.

% even if a topologically complicated and active magnetic field is responsible for the FRB activity. 

\begin{figure}
%\centering
\includegraphics[width=0.473\textwidth]{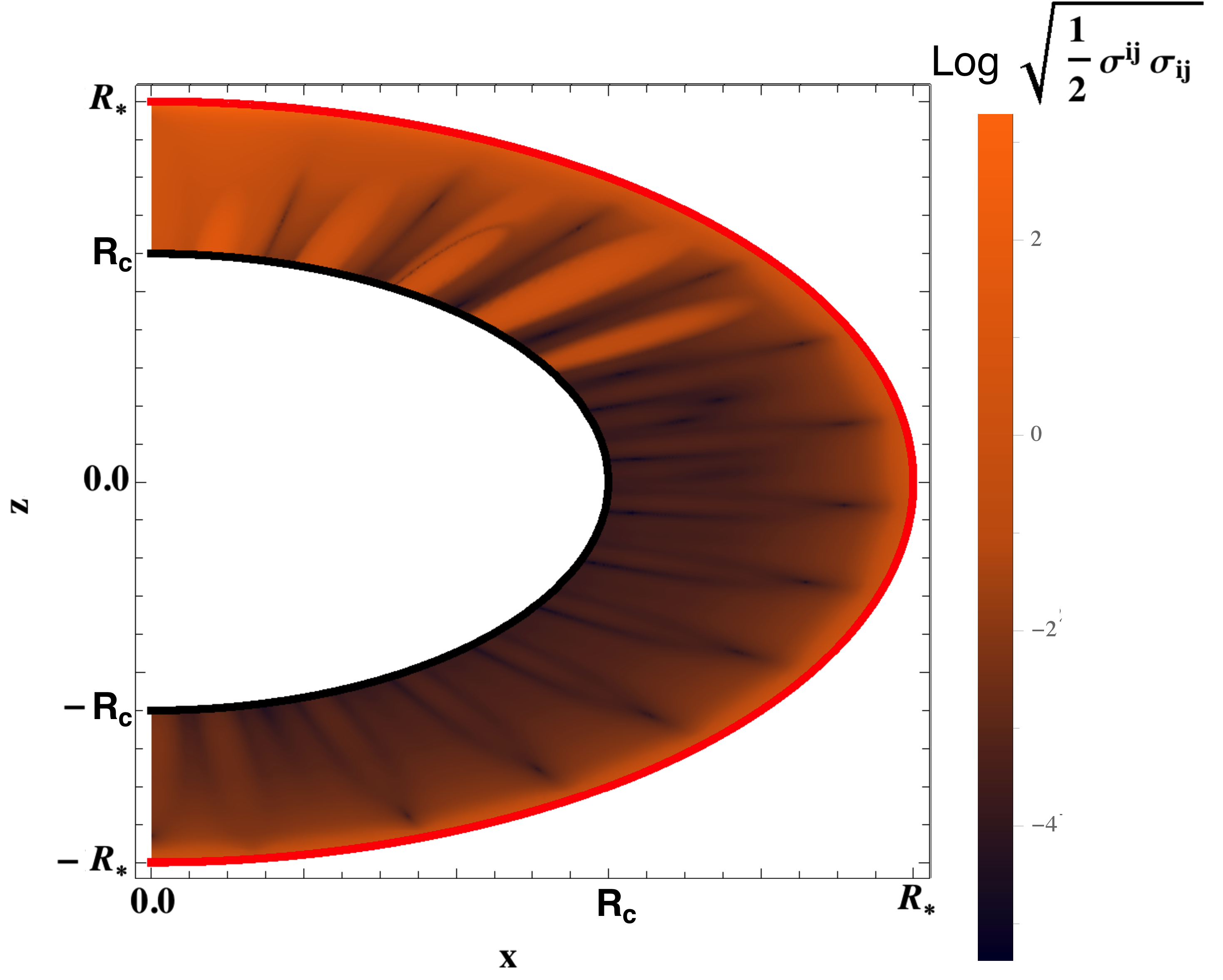}
\caption{Similar to Fig. \ref{sigmaa} though for the field configurations defined within model B. \label{sigmab}
}
\end{figure}

\begin{figure}
%\centering
\includegraphics[width=0.473\textwidth]{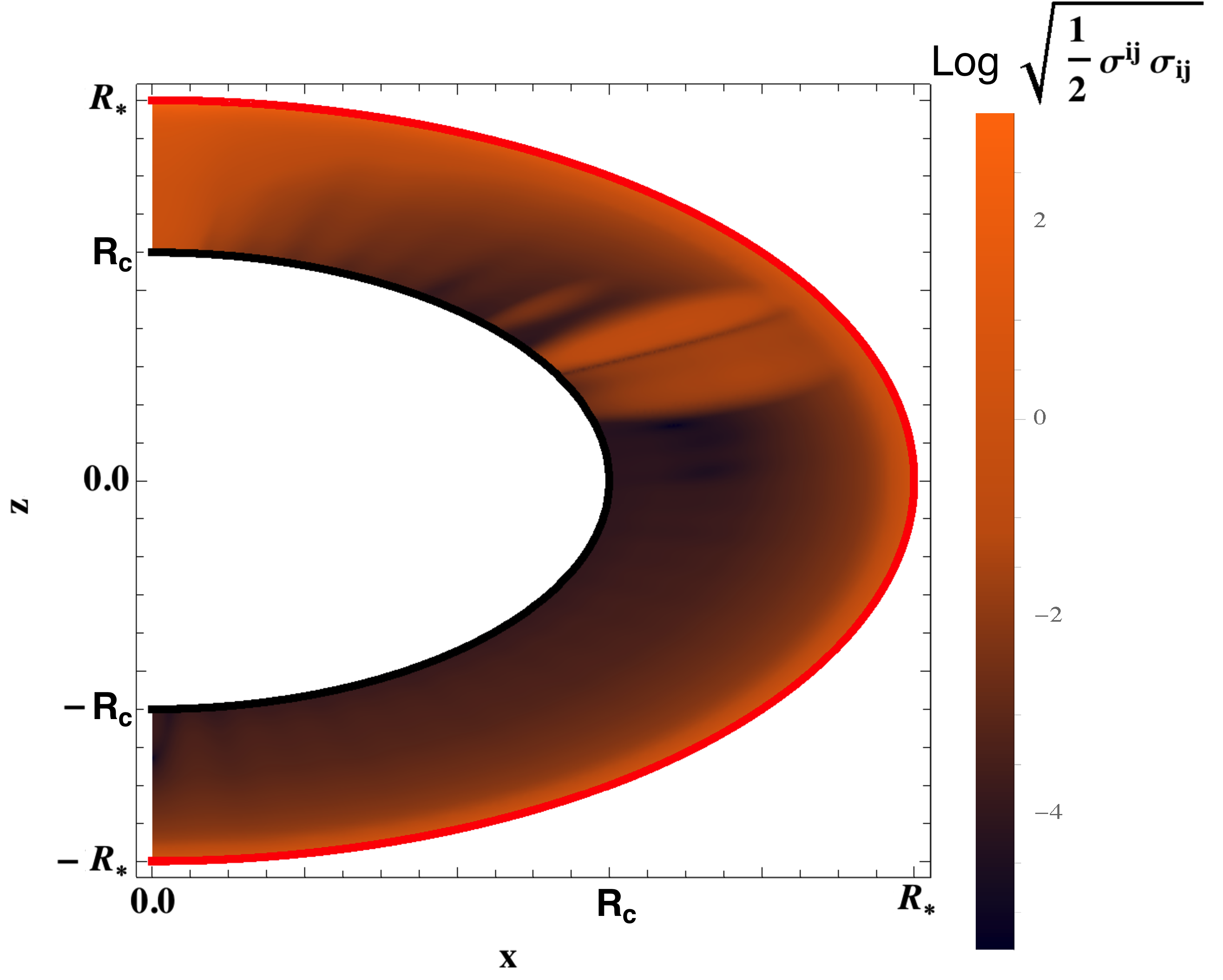}
\caption{Similar to Fig. \ref{sigmaa} though for the field configurations defined within model C. \label{sigmac}
}
\end{figure}

Figures \ref{sigmab} and \ref{sigmac} depict the contracted strain tensor for models B and C, respectively. In both Figs. \ref{sigmab} and \ref{sigmac}, we see that the fracture geometry follows the multipolar `plumes' as before. In Fig. \ref{sigmac}, which shows the strain from model C, which has a final state where the dipole component is strong, the magnetic walls are especially apparent in the southern hemisphere, and the fracture geometry is topologically complicated. In particular, many deep $(d \lesssim R_{c})$ but short $(l \lesssim 10^{3} \text{ cm})$ quakes may occur from crust yielding in this scenario. %While in Fig. \ref{sigmab} for model B, the fracture geometry is

%This situation may represent an epoch in which the star is particularly active from an FRB perspective. %as it grows older and the dipole field begins to become dominant. %While in Fig. \ref{sigmab}, where the fields are strongly multipolar in both cases $(\xi \approx 0)$, the star may be more prone to higher energy, but less frequent, FRB activity.

The fact that each of the multipolar components have different amounts of energy stored within them implies that the Hall timescale \eqref{eq:hall} differs slightly for each $\ell$. The implication is that the energetics associated with the fracture geometry through \eqref{eq:vonb} and the waiting times may then be seemingly uncorrelated [depending on $\xi(t)$], consistent with the statistical findings of \cite{li19} for FRB 121102.

Between each of the models presented, we have that the toroidal field has decayed slightly $(\Lambda_{f} > \Lambda_{i})$, thereby causing an energy differential of $\approx 10^{45} \text{ erg}$ between the initial and final states. The Weibull distribution analysis of \cite{opp18} suggests that FRB 121102 may have a mean repetition rate of $\approx 5 \pm 2 $ days\footnote{Recently, \cite{zhang18b} reported the discovery of $93$ pulses in FRB 121102 within $5$ hours. The quoted mean repetition rate of \cite{opp18} may therefore be an underestimate.}, thereby implying that, over $\sim 10^{2}$ years, a total of $\sim 10^{4}$ bursts may occur. If each of these contains roughly $\sim 10^{39} \text{ erg}$ of energy in accord with \eqref{eq:frbenergy}, this would suggest that the total energy released over the bulk Hall time \eqref{eq:hall} is $\sim 10^{43} \text{ erg}$. This is consistent with the quake model here, provided that $\approx 1\%$ of the lost magnetic energy is converted into FRBs \citep{wad19}.  

%HAVE SOME ACTUAL `COUNT' OF CRUSTAL FRACTURES, AND HOW HOW MANY BURSTS OR SOMETHING? MAYBE HARD. CAN USE THEIR RELATIVE DEPTHS FROM THE PLOTS TO SEE THOUGH.

\section{Discussion}

%We have  shown that it is easily possible for a young magnetar with a complicated magnetic field to unleash seemingly uncorrelated

%Furthermore, if young magnetars fracturing crusts are responsible for repeating FRBs, one expects the natal magnetic field to be large, as a

In this paper, inspired by the statistical findings of \cite{wang18}, which suggest a crustquake progenitor for repeating FRBs (Sec. 2), we have investigated how the evolution of a magnetic field within the crust of a young magnetar can be related to FRB properties (Sec. 3). By adopting models which, while simple, encapsulate the major features of the sophisticated numerical simulations of \cite{gourg16}, we have shown how quake geometry \eqref{eq:vonb}, dictated by the magnetic field \eqref{eq:bfield}, can be related to the burst energetics \eqref{eq:equake} (Sec. 4). If a magnetar is born with a sufficiently strong, tangled magnetic field, predicted to occur under some circumstances following core-collapse \citep{ober17,ober18} and NS-NS merger events \citep{giac15,ciolfi19}, the Hall time \eqref{eq:hall} can be sufficiently short so as to instigate rapid field evolution in the crust over $\lesssim 10^{2}$ yr [in agreement with the predicted age of the object within FRB 121102 \citep{bower17,metzger17}], generating magnetic stresses which crack the crust and release energy. We find that a multipolar magnetic field is important in the scenario because it allows for multiple, isolated fractures, each of which contribute separately to the overall burst activity, alleviating the concerns of \cite{li19} concerning the waiting time statistics of FRB 121102.
% locations over the course of the Hall time \eqref{eq:hall}.

If young magnetars with especially strong crustal fields $B \gtrsim 10^{15} \text{ G}$ are responsible for repeating FRBs, one would expect that the birth rate of newborn magnetars, with the requisite characteristics, can be matched with the number of observed sources. Statistical studies of type II, Ib, and Ic supernovae suggest that approximately $1.5 \pm 1$ NSs are born within our galaxy every century \citep{tamm94,janka04,diehl06}. If we assume that this statistic is roughly constant amongst galaxies at low redshift, we can get an estimate for the number of neutron stars born per year within a certain distance, e.g. within $D \lesssim 1$ Gpc [i.e. the distance to FRB 121102 \citep{chatt17}]. From distributions of dark matter halos at low redshift \citep{murray13}, one can estimate, assuming that each halo hosts a galaxy, that $10^{7} \lesssim N_{\text{gal}}(z \leq 0.2) \lesssim 10^{9}$, where $N_{\text{gal}}$ is the number of (generic) galaxies. While it is generally expected that $\approx 10\%$ of NSs are born as soft-gamma repeaters or anomalous X-ray pulsars \citep{kouv94,muno07} [though cf. \cite{ben19} who claim that it may be as much as $40 \%$], only a small fraction of these are expected to house the strongest magnetic fields $B > 10^{15} \text{ G}$. The population synthesis models of \cite{popov10} suggest\footnote{Note that \cite{popov10} consider field strengths characteristic of the \emph{whole} star, including the core. As such, the percentage of magnetars with \emph{crustal} field strengths of this order is likely to be even lower, since the core field may be $\gtrsim 10$ times stronger than the crustal field.} that as few as $1$ in $\sim 10^{5}$ of newborn NSs may have magnetic field strengths $B \geq 3 \times 10^{15} \text{ G}$. Therefore, the number of magnetars, $N_{\text{mag}}$, with field strength $B \gtrsim 10^{15} \text {G}$ born within $\sim 1$ Gpc can be estimated as
\begin{equation} \label{eq:magnetarrate}
\frac {10^{2}}  {10^{2} \text{ yr} } \lesssim \dot{N}_{\text{mag}}(z \lesssim 0.2)  \lesssim \frac {10^{5}}  {10^{2} \text{ yr} }.
\end{equation}
Note that expression \eqref{eq:magnetarrate} estimates the number of potentially active sources within $\sim 1$ Gpc. While, even on the lower end, \eqref{eq:magnetarrate} is still considerably greater than $2$, when considering the \emph{observable} number of repeating FRB sources, there are other factors to consider. In particular, taking into account reductions related to the beaming fraction (i.e. the fraction whose emissions pass by Earth), the possibility of sources having dormant and active epochs \citep{zhang18b,price18}, that field strength is not the only factor (i.e. highly multipolar fields might only occur within some fraction of those which have strong fields), the number of observable sources will be much lower than the estimate in \eqref{eq:magnetarrate}. Though, a more careful analysis of the FRB emission mechanism, NS orientations, and magnetar formation rates in low redshift galaxies is necessary to make a more definitive conclusion. %which is beyond the scope of this paper.,mlm13,msm15

In general, because magnetic fields deform a star and induce a mass quadrupole moment, millisecond magnetars are expected to be excellent sources of gravitational waves \citep{dall09,mmra11,dall15}. Unfortunately, owing to the $\lesssim$ Gpc distances to the repeating sources FRB 121102 and FRB 180814.J0422+73, it is highly unlikely that gravitational wave counterparts will be found since the signal-to-noise ratio $S/N$ scales as $D^{-1}$. Furthermore, since the (electromagnetic) spindown timescale is short for a magnetar, $t_{\text{sd}} \sim 2 \times 10^{3} \left( B / 10^{15} \text{ G} \right)^{-2} \left( P / \text{ ms} \right)^{2}$ s \citep{lu18}, older sources are harder to detect since the gravitational wave strain $h \propto P^{-2}$. 

However, the strain $h$ is also proportional to the square of the magnetic field strength through the gravitational ellipticity $\epsilon$, so a NS with a strong $(B \gtrsim 10^{15} \text {G})$ and tangled magnetic field within $\lesssim 20$ Mpc of Earth would likely be detectable within the early stages of its life, with facilities such as the Laser Interferometer Gravitational-Wave Observatory (LIGO) or the upcoming Einstein telescope, with $S/N \gg 10$ \citep{dall09,dall15}. The detectability increases further if the star houses a strong, toroidal field $(\Lambda \ll 1)$ \citep{cut02,mlm13,msm15,suv16}. {Using the non-barotropic approach of \cite{mlm13,msm15}, the perturbed density associated to a stellar magnetic field is estimated through}
 \begin{equation} \label{eq:deltarho}
 \frac{ \partial \delta \rho} {\partial \theta} = - \frac {r} { 4 \pi R_{\star}} \left( \frac {d \Phi} {d r} \right)^{-1} \left\{ \nabla \times \left[ \left( \nabla \times \boldsymbol{B} \right) \times \boldsymbol{B}  \right] \right\}_{\phi},
 \end{equation}
{where $\Phi$ is the gravitational potential for the equilibrium configuration. From \eqref{eq:deltarho}, the gravitational ellipticity $\epsilon$ can be found, viz.}
 \begin{equation} \label{eq:epsilon}
 \epsilon = \pi I_{0}^{-1} \int_{V} dr d \theta \delta \rho(r, \theta) r^4 \sin \theta \left( 1 - 3 \cos^2\theta \right),
 \end{equation}
{for moment of inertia $I_{0}$. In Table \ref{tab:ellipdata}, we list the ellipticities for the initial and final states of models A, B, and C. In all cases except the initial state for model B, the toroidal field is strong enough to deform the star into a prolate shape $(\epsilon < 0)$. In general, we find that the oblateness contributed by the poloidal field is of the order $\epsilon \lesssim 10^{-5}$, in agreement with other estimates found in the literature \citep{dall09,dall15,suv16}. It is generally expected that the wobble angle $\vartheta$ of a precessing, prolate star with misaligned magnetic and angular momentum axes tends to grow until $\vartheta = \pi/2$, which is the optimal state for gravitational wave emission \citep{cut02}. Significant gravitational radiation and freebody precession would therefore be expected from the models (C especially) presented here \citep{jones02,gual10}.} If, in the future, a repeating FRB source is detected within (say) the Virgo cluster, a coincident measurement of gravitational waves would place strong constraints on the nature of the object, however this would likely disfavour the (Hall-driven) quake scenario because of the disparity between the Hall time \eqref{eq:hall} and the spin-down time, $\tH/t_{\text{sd}} \sim 10^{6}$.

%Thus, since $|\epsilon|$ is large for the models presented here (model C in particular), a detection of gravitational waves may be easier than for an oblate star, with equal $|\epsilon|$, as the wobble angle is rapidly damped for oblate configurations \citep{jones02}.}
\begin{table*}
\caption{Ellipticites \eqref{eq:epsilon} of the initial and final states of models A, B, and C, detailed in Sec. 4.3, computed assuming a parabolic (Tolman VII) equilibrium density profile, $\rho = \tfrac{15 M_{\star}}  { 8 \pi R_{\star}^3 } \left( 1 - \tfrac{r^2}{R_{\star}^2} \right)$ with $M_{\star} = 1.4 M_{\odot}$ and $R_{\star} = 10^{6} \text{ cm}$, to determine $\Phi(r)$ from the Poisson equation $\nabla^2 \Phi = 4 \pi G \rho$.
}
  \begin{tabular}{llcccc}
  \hline
Model & $\epsilon_{i}$ $(\times 10^{-5})$ & $\epsilon_{f}$ $(\times 10^{-5})$ \\
\hline
A & $-5.06$ & $-4.44$ \\
B & $0.177$ & $-0.515$ \\
C & $-10.0$ & $-6.78$ \\
\hline
\end{tabular}
\label{tab:ellipdata}
\end{table*}

{Finally, it is important to note that we have not included effects related to plastic flow, such as those discussed by \cite{lan19}. In particular, Hall-induced magnetic stresses shear the crust and may initiate a plastic flow, as opposed to fracturing, which dissipatively converts excess stress beyond the critical threshold (in the von Mises sense) to heat at a rate which depends on the plastic viscosity, reducing the overall stress felt by the outer layers of the star \citep{jones03,belo14}. Plastic flow is also known to induce a logarithmic creep, which tends to postpone subsequent quakes \citep{baym71}. Plastic flow effects may therefore limit the ability of the star to release energy over the timescales required for repeated FRB activity. However, some of the heat deposited due to the plastic flow is eventually conducted to the stellar surface, possibly initiating afterglow activity within $\sim$ yrs after a burst \citep{li15}, which may be connected to magnetar activity and have observable consequences. A thorough investigation of such effects requires full (general relativistic) elastic magnetohydrodynamics simulations using, for example, the formalism developed by \cite{and19}.}

% For example, no detection of burst radiation from an FRB repeater within $\lesssim 20$ Mpc would disfavour a magnetar progenitor \citep{dall15}. %using facilities such as the Laser Interferometer Gravitational-Wave Observatory (LIGO).

%, the magnetic field required to instigate the crustal fractures would easily result in gravitational radiation that should be detectable with $S/N \gg 10$  \citep{dall09,mmra11,suv16}, especially for stars housing strong, toroidal fields \citep{cut02}.

%The energetics associated here may be related to other events such as magnetar flares \citep{col11,zink12} or (anti-)glitch activity \citep{eps00,mast15}

%As aquick example, consider a newborn magnetar in the Virgo cluster, rotating with initial spin period 0.97 ms and final spinperiod 10 s, withB0= 5×1010T (Dall’Osso, Shore, & Stella 2009; Mastrano et al. 2011). This magnetar has significantdetectability (S/N >10) for .

%A detection of a young ($\lesssim 10^{2}$ year old) magnetar with strong magnetic field in the absence of FRBs would cast doubt on our model.

\section*{Acknowledgements}
{We thank the anonymous referee for their insights and helpful suggestions.} This work was supported by the Alexander von Humboldt Foundation.

\bsp \label{lastpage}

\end{document}